
%
%
%
%
\def\@{{\char'100}}

\long\def\abstract#1{\bigskip{\advance\leftskip by 2true cm
\advance\rightskip by 2true cm\eightpoint\centerline{\bf
Abstract}\everymath{\scriptstyle}\vskip10pt\vbox{#1}}\bigskip}
\long\def\resume#1{{\advance\leftskip by 2true cm
\advance\rightskip by 2true cm\eightpoint\centerline{\bf
R\'esum\'e}\everymath{\scriptstyle}\vskip10pt \vbox{#1}}}
\def\smallmatrix#1{\left(\,\vcenter{\baselineskip9pt\mathsurround=0pt
\ialign{\hfil$\scriptstyle{##}$\hfil&&\ \hfil$\scriptstyle{##}$\hfil
\crcr#1\crcr}}\,\right)}
\def\references{\bigbreak\centerline{\sc
References}\medskip\nobreak\bgroup
\def\ref##1&{\leavevmode\hangindent45pt
\hbox to 42pt{\hss\bf[##1]\ }\ignorespaces}
\parindent=0pt
\everypar={\ref}\par}
\def\endreferences{\egroup}

\long\def\authoraddr#1{\medskip{\baselineskip9pt\let\\=\cr
\halign{\line{\hfil{\Addressfont##}\hfil}\crcr#1\crcr}}}
\def\Untertitel#1{\medbreak\noindent{\Untertitelfont#1.} }
%
%
\newif\ifrunningheads
\runningheadstrue
\immediate\write16{- Tetes de page}
\headline={\ifrunningheads\ifnum\pageno=1\hfil\else\ifodd\pageno\rightheadline
\else\leftheadline\fi\fi\else\hfil\fi}
\def\rightheadline{\sc\hfil\RightHeadText\hfil}
\def\leftheadline{\sc\hfil\LeftHeadText\hfil}

%
%
%
\def\operatorname#1{{\rm#1}}
\def\tr{\operatorname{tr}}
\def\res{\operatorname{res}}
\def\Ad{\operatorname{Ad}}
%
%
\immediate\write16{- Caracteres "Small Caps" et "Euler
Fraktur"}
%
%
%

\let\sc=\tensmc
%
%
\font\teneuf=eufm10  \font\seveneuf=eufm7 \font\fiveeuf=eufm5
\newfam\euffam \def\gr{\fam\euffam\teneuf}
\textfont\euffam=\teneuf \scriptfont\euffam=\seveneuf
\scriptscriptfont\euffam=\fiveeuf
%

	\def\grg{{\gr g}}

	\def\grl{{\gr l}}

	\def\grs{{\gr s}}


\def\sig{\sigma}		

\def\ome{\omega}		\def\Ome{\Omega}
\def\nchi{\hbox{\raise 2.5pt\hbox{$\chi$}}}


\def\CalD{{\cal D}}

\def\CalH{{\cal H}}
\def\CalI{{\cal I}}

\def\CalN{{\cal N}}
\def\CalO{{\cal O}}

%

\def\bfg{{\bf g}}		
		
		\def\bfI{{\bf I}}

		\def\bfR{{\bf R}}

\def\Mtil{{\widetilde M}}
\def\util{{\tilde u}}
%
%
\def\authorfont{\sc}
\font\eightrm=cmr8
\font\eightbf=cmbx8
\font\eightit=cmti8
\font\eightsl=cmsl8

\def\eightpoint{\let\rm=\eightrm \let\bf=\eightbf \let\it=\eightit
\let\sl=\eightsl \baselineskip = 9.5pt minus .75pt  \rm}

\font\titlefont=cmbx10 scaled\magstep2
\font\sectionfont=cmbx10
\font\Untertitelfont=cmbxsl10
\font\Addressfont=cmsl8
%
%
\def\Satz#1:#2\par{\smallbreak\noindent{\sc #1:\ }
{\sl #2}\par\smallbreak}
%
%
\immediate\write16{- Debuts de chapitres.}
\newcount\secount
\secount=0
\newcount\eqcount
\outer\def\section#1.#2\par{\global\eqcount=0\bigbreak
\ifcat#10
 \secount=#1\noindent{\sectionfont#1. #2}
\else
 \advance\secount by 1\noindent{\sectionfont\number\secount. #2}
\fi\par\nobreak\medskip}
%
%
\immediate\write16{- Numerotation automatique.}
\catcode`\@=11
\def\adv@nce{\global\advance\eqcount by 1}
\def\unadv@nce{\global\advance\eqcount by -1}
\def\nextnumber{\adv@nce}
%
%
\newif\iflines
\newif\ifm@resection
\def\onesec{\m@resectionfalse}
\def\moresec{\m@resectiontrue}
\moresec
\def\eq{\global\linesfalse\eq@}
\def\eqn{\global\linestrue&\eq@}
\def\nosubind@x{\global\subind@xfalse}
\def\newsubind@x{\ifsubind@x\unadv@nce\else\global\subind@xtrue\fi}
\newif\ifsubind@x
\def\eq@#1.#2.{\adv@nce
 \if\relax#2\relax
  \edef\loc@lnumber{\ifm@resection\number\secount.\fi
  \number\eqcount}
  \nosubind@x
 \else
  \newsubind@x
  \edef\loc@lnumber{\ifm@resection\number\secount.\fi
  \number\eqcount#2}
 \fi
 \if\relax#1\relax
 \else
  \expandafter\xdef\csname #1@\endcsname{{\rm(\loc@lnumber)}}
  \expandafter
  \gdef\csname #1\endcsname##1{\csname #1@\endcsname
  \ifcat##1a\relax\space
  \else
   \ifcat\noexpand##1\noexpand\relax\space
   \else
    \ifx##1$\space
    \else
     \if##1(\space
     \fi
    \fi
   \fi
  \fi##1}\relax
 \fi
 \eq@@{\loc@lnumber}}
\def\eq@@#1{\iflines \else \eqno\fi{\rm(#1)}}
\def\m@th{\mathsurround=0pt}
%
%
\def\display#1{\null\,\vcenter{\openup1\jot
\m@th
\ialign{\strut\hfil$\displaystyle{##}$\hfil\crcr#1\crcr}}
\,}
\newif\ifdt@p
\def\@lign{\tabskip=0pt\everycr={}}
\def\displ@y{\global\dt@ptrue \openup1 \jot \m@th
 \everycr{\noalign{\ifdt@p \global\dt@pfalse
  \vskip-\lineskiplimit \vskip\normallineskiplimit
  \else \penalty\interdisplaylinepenalty \fi}}}
%
%
\def\displayno#1{\displ@y \tabskip=\centering
 \halign to\displaywidth{\hfil$
\@lign\displaystyle{##}$\hfil\tabskip=\centering&
\hfil{$\@lign##$}\tabskip=0pt\crcr#1\crcr}}
%
%
\def\cite#1{{\bf[#1]}}
\catcode`\@=\active
\magnification=\magstep1
\hsize= 6.75 true in
\vsize= 8.75 true in
%
%
%
\def\la{\lambda}

\def\PC{{\bf P^1(C)}}

\def\ra{\rightarrow}
\def\gR{{\tilde\grg_R}}
\def\sym{\ome=dx_1\wedge dy_1+dx_2\wedge dy_2}
\def\RightHeadText{Hamiltonian Models for Painlev\'e Transcendants}
\def\LeftHeadText{J. Harnad and M.A. Wisse}
%
%
\rightline{CRM-1878 (1993) \break}
\bigskip
\centerline{\titlefont Loop Algebra Moment Maps and Hamiltonian Models}
\centerline{\titlefont for the Painlev\'e Transcendants}
\bigskip
\centerline{\authorfont J.~Harnad\footnote{$^{\scriptstyle1}$}{\eightpoint
e-mail address: harnad\@alcor.concordia.ca {\it\ or\ }
harnad\@mathcn.umontreal.ca}}
\authoraddr
{Department of Mathematics and Statistics, Concordia University\\
7141, Sherbrooke W., Montr\'eal, Qu\'ebec, Canada H4B 1R6, and\\
Centre de recherches math\'ematiques, Unversit\'e de Montr\'eal\\
C.~P.~6128, succ.``A'', Montr\'eal, Qu\'ebec, Canada H3C 3J7}
\bigskip
\centerline{\authorfont M.~A.~Wisse\footnote{$^{\scriptstyle2}$}
{\eightpoint e-mail address: wisse\@mathcn.umontreal.ca}}
\authoraddr
{D\'epartement de math\'ematiques et de statistique, Universit\'e de
Montr\'eal\\
C.~P.~6128, succ."A", Montr\'eal, Qu\'ebec, Canada H3C 3J7}

\abstract{The isomonodromic deformations underlying the Painlev\'e
transcendants are interpreted as nonautonomous Hamiltonian systems in the
dual $\gR^*$ of a loop algebra $\tilde\grg$ in the classical $R$-matrix
framework. It is shown how canonical coordinates on symplectic vector spaces
of dimensions four or six parametrize certain rational coadjoint orbits in
$\gR^*$ via a moment map embedding. The Hamiltonians underlying the Painlev\'e
transcendants are obtained by pulling back elements of the ring of
spectral invariants. These are shown to determine simple Hamiltonian systems
within the underlying symplectic vector space.}

\bigskip
\line{\hfill\sl Dedicated to Jerrold Marsden on the occasion of his
 50th birthday. \hfill}

\bigbreak
\noindent{\sectionfont Introduction}\medskip\nobreak

The Painlev\'e transcendants $P_I-P_{VI}$ have been interpreted as
isomonodromic
deformation equations for first order matrix systems $\partial\psi/\partial
\la=\CalN(\la)\psi$, where $\CalN(\la)$ is rational in $\la$, in \cite{FN, JM}.
The Hamiltonian structure of these equations has been studied by many authors
(see e.g. \cite{Ok} and references therein).
In this paper, the Hamiltonians for $P_I-P_{VI}$ are derived within the
framework of the Adler-Kostant-Symes (AKS) theorem and the classical
$R$-matrix approach \cite{S} on a rational coadjoint orbit in the dual
$\tilde\grg^{*}$ of a loop algebra $\tilde\grg$. This construction allows for
the parametrization of the underlying phase space by Darboux coordinates on a
finite-dimensional symplectic vector space $M$ via a moment map embedding,
according to a general scheme developed in \cite{AHP, AHH1, AHH2}. The AKS
Hamiltonians for $P_I-P_{VI}$ on $\tilde\grg^*$ pull back to $M$ where they may
be interpreted as simple mechanical systems. In each case the Painlev\'e
transcendant is obtained by Hamiltonian reduction. This scheme is used in
\cite{H} to parametrize the phase spaces for $P_V$ and $P_{VI}$. In \cite{HTW}
a special case of $P_{II}$ (and $P_{IV}$) arising in the spectral theory of
random matrices is given a similar loop algebra formulation. In the present
work we give a complete list of Hamiltonian
systems related to the Painlev\'e transcendants obtained within this
framework.

\section 1.Darboux Coordinates on Rational Coadjoint Orbits in Loop Algebras.

\Untertitel{1a. Loop algebras}Let $\grg$ be a Lie algebra. For the purpose of
this paper it is sufficient to consider $\grg=gl(2)$. Let ${\cal C}$ be a
closed, simple curve in $\PC$ (typically, a circle
centered at the origin), dividing it into an interior region
$U^+$,
containing $0$ and an exterior region
$U^-$, containing $\infty$. The space $\tilde\grg$ of smooth maps
$X:{\cal C}\rightarrow\grg$ with the natural Lie algebra structure induced by
that on $\grg$ defines our loop algebra. We consider the
splitting of $\tilde\grg$ as a vector space direct sum
$$
\tilde\grg=\grg^+\oplus\grg^-\eq..
$$
of the subalgebra $\grg^+$, consisting of maps admitting a holomorphic
extension to $U^+$ and $\grg^-$, consisting of maps $X$ admitting a
holomorphic extension to $U^-$, the latter normalized by the condition
$X(\infty)=0$. On $\tilde\grg$ we define the Ad-invariant inner product
$$
<X,Y>={1\over 2\pi i}\oint_{\cal C}\tr(X(\la)Y(\la))d\la,\eq inner..
$$
which provides an identification of $\tilde\grg$ with a dense subspace of
$\tilde\grg^{*}$. In the dual decomposition
$$
\tilde\grg^*=\grg^{+*}\oplus\grg^{-*},\eq..
$$
we have the corresponding identifications of $\grg^\pm$ with
dense subspaces of $(\grg^\mp)^*$.

Let $P_\pm:\tilde\grg\ra\grg^\pm$ be the projections to
these subalgebras, and define
$$
R:={1\over2}(P_+-P_-).\eq..
$$
Then the bracket defined by
$$
\lbrack X,Y\rbrack_R=\lbrack R(X),Y\rbrack + \lbrack X,R(Y) \rbrack\eq
Rmat..
$$
determines a second Lie algebra structure on $\tilde\grg$, and $R$ is referred
to
as a ``classical $R$-matrix''. We shall write $\tilde\grg_R$ when referring to
$\tilde\grg$ endowed with the Lie bracket \Rmat and
denote by $\Ad^*_R$ the coadjoint representation with respect to this modified
Lie bracket, while $\Ad^*$ continues to denote the ordinary coadjoint
representation.

On $\gR^*$ we have the Lie-Poisson
structure given by
$$
\{f,g\}_R(\mu)=\mu(\lbrack df(\mu),dg(\mu)\rbrack_R),\ f,g\in
C^\infty(\gR^*), \mu\in\gR^*.\eq..
$$
(The Lie Poisson structure with respect to the original Lie bracket on
$\tilde\grg$ plays no direct role in what follows.)

\Untertitel{1b. Moment maps into loop algebras}Let $M=\bfR^{2N}$ be a
finite-dimensional symplectic vector space with symplectic
form $\Ome$ given by
$$
\Ome=\sum_{i=1}^n dx_i\wedge dy_i.\eq..
$$
The Poisson space $\gR^*$ contains finite dimensional Poisson subspaces with
symplectic leaves consisting of orbits whose elements are rational in $\la$.
Let
$\bfg_A\subset\gR^*$ be such a finite-dimensional Poisson-subspace. A
Poisson map
$$
J:M\ra\bfg_A\subset\gR^*\eq MM..
$$
whose coefficients generate complete flows may be interpreted as an equivariant
moment map for a suitable group action. A
general scheme for the construction of such moment maps into loop algebras with
applications to ODE's and PDE's can be found in \cite{AHP, AHH1, AHH2,
HW}.

The image of a moment map of type \MM consists of elements of the form
$$
\CalN(\la)=\la^{n_0}Y+\sum_{l_0=0}^{n_0-1}N_{0,l_0}\la^{l_0}+\sum_{i=1}^n
\sum_{l_i=1}^{n_i}{N_{i,l_i}\over(\la-\alpha_i)^{l_i}},\eq PM..
$$
where the poles $\alpha_i$ are viewed as parameters characterizing the
$\Ad_R^*$-orbits, $Y\in\grg$ is a constant element whose components are
Casimir invariants on $\bfg_A$ and the entries of $N_{k,l}\in\grg$ are
typically rational expressions in $(x_i,y_i)$.

In the examples below it will be possible to
choose $J$ such that it is locally a symplectic diffeomorphism onto a coadjoint
orbit $\CalO$ in $\bfg_A$, thereby parametrizing $\CalO$ in terms of the
Darboux coordinates on $M$.

\Untertitel{1c. Autonomous and nonautonomous Hamiltonian systems in loop
algebras and isomonodromic deformation equations}Let
${\cal I}$ be the ring of $\Ad^*$-invariant polynomial functions
on $\tilde\grg^*_R$, restricted to $\bfg_A$. These functions will also be
called {\it spectral invariants}. The
Adler-Kostant-Symes (AKS) theorem, in its generalized $R$-matrix form,
\cite{S} tells us that
\item{(1)} elements of ${\cal I}$ Poisson-commute (and
hence so do their pullbacks under the Poisson map $J$).
\item{(2)} Hamilton's equations for $H\in{\cal I}$ are
given by
$$
{d{\cal N}\over dt}=\lbrack P_\sig\circ dH({\cal
N}),{\cal N}\rbrack,\quad  {\cal N}\in\bfg_A,\eq
Hameqs..
$$
where
$$
P_\sig=(1+\sig)P_+ +\sig P_-,\quad \sig\in \bfR.\eq..
$$
(The equivalence of this equation for all values of $\sig\in\bfR$ follows from
the $\Ad^*$-invariance of $H$.)

\noindent
These autonomous systems therefore preserve the spectrum of
$\CalN(\la)$, and the ring of invariants is identified with the
coefficients of the characteristic equation
$$
\det(\CalN(\la)-\zeta\bfI)=0\eq..
$$
defining the ``spectral curve'' $X$. Identifying one or more of the
$\alpha_i$'s or the entries of $Y$ with the flow-parameter $t$, the
resulting nonautonomous system is
given by
$$
{\partial{\cal N}\over \partial t}=\lbrack P_\sig\circ dH({\cal N}),{\cal
N}\rbrack+{\cal N}_t,\quad{\cal N}\in\bfg_A,\eq naH..
$$
where $\CalN_t$ denotes the partial derivative with respect to the explicit
dependance on $t$. This may still be viewed as a Hamiltonian system on $M$,
induced by the nonautonomous Hamiltonian $J^*(H)$. The system \naH is no longer
isospectral, but defines an isomonodromic deformation equation if the
condition
$$
{\partial{\cal N}\over\partial t}={\partial\over\partial\la}
(P_\sig\circ dH({\cal N}))\eq..
$$
is satisfied.
The nonautonomous system \naH is then equivalent to the
commutativity
$$
\lbrack \CalD_\la,\CalD_t\rbrack=0\eq..
$$
of the differential operators
$$
\eqalignno{
{\cal D}_\la=&{\partial\over\partial\la}-{\cal N}(\la)\eqn Dopsa.a.\cr
{\cal D}_t=&{\partial\over\partial t}-P_\sig\circ dH({\CalN})(\la).
\eqn Dopsb.b.}
$$
An important class of Hamiltonian systems on $\bfg_A$ leading to systems
of differential operators defining isomonodromic deformation equations is
obtained if $Y$ is diagonalizable, $n_0=0$ and the poles in \PM are simple.
(See \cite{H} for details).

\Untertitel{1d. Reductions}From now on we specialize to the case $r=2$ in order
to simplify the exposition. This is the case relevant for the Painlev\'e
transcendants. The elements of $\CalI$ are invariant under
the action of the (generically 1-dimensional) isotropy group $H_Y\subset Gl(2)$
of $Y$ (quotienting by the centre to obtain an effective
action). It follows that the spectral invariants project to the
Marsden-Weinstein
reduction of $\CalO$ by $H_Y$. These reductions will be shown to give the phase
spaces underlying the
Painlev\'e transcendants $P_{I}-P_{V}$. For $P_{VI}$ the image of $J$ is
entirely included in $\tilde\grg^{+*}$, so that $Y=0$, and therefore
$H_Y=Sl(2)$. In this case the ``unreduced'' phase space is obtained by
partially fixing the moment map $J_{H_Y}$ generating the $Sl(2)$-action. The
corresponding level sets will be chosen so as to define a symplectic
submanifold
of $\CalO$.

\Untertitel{1e. Canonical spectral coordinates}A coadjoint orbit consisting of
elements of the form \PM may alternatively be parametrized by ``spectral
Darboux
coordinates'' $(\ln(w),a,u_i,v_i)$, $i=1,\dots,g$, where $g$ is the
genus of the spectral curve $X$. In the autonomous,
isospectral case, these lead to a complete separation  of variables on
generic coadjoint orbits, and thereby a linearization of the flows of
the Jacobi variety of $X$ (cf. \cite{AHH3}).
Taking $Y_{21}=0$ they are defined by the equations
$$
\eqalign{
\CalN(\la)_{21}&={w\prod_{i=0}^g(\la-u_i)\over \tilde{a}(\la)},\cr
\CalN(u_i)_{22}&=v_i,}\eq spcoordeqs..
$$
and $\tilde{a}(\la)=\prod_{i=1}^n(\la-\alpha_i)^{n_i}$. To complete the system
we must add the generator $a$ of the 1-dimensional group $H_Y$, which is
canonically conjugate to the ``ignorable'' coordinate $w$. The resulting
symplectic form on $\CalO$, before the reduction by $H_Y$, is given by
$$
\ome=\sum_{i=1}^g du_i\wedge dv_i+d\ln(w)\wedge da.\eq..
$$
(In certain cases, it is convenient to make a further canonical coordinate
transformation $v_i\ra v_i+f_i(u_i)$ to simplify the resulting expressions.)
Choosing a level set for $a$; that is, performing the Marsden-Weinstein
reduction  by $H_Y$, we obtain a symplectic manifold with reduced symplectic
form
$$
\ome_{\rm red}=\sum_{i=1}^g du_i\wedge dv_i.\eq..
$$

For the examples given below, we always have $g=1$, and $u_1=:u$ is,
essentially, the variable satisfying the Painlev\'e equation.

\section.Hamiltonian Models for the Painlev\'e Transcendants.

\Untertitel{Painlev\'e I}
Let $M=\bfR^4$, with symplectic form
$$
\sym.\eq ddimsym..
$$
The moment map $J_I:\bfR^4\ra\gR^*$ is defined by
$$
J_I(x_1,x_2,y_1,y_2):=\CalN(\la)=\la^2 \pmatrix{0&1\cr0&0}+\la\pmatrix{x_1 &
x_2
\cr \kappa & -x_1} +\pmatrix{-y_2+x_1 x_2 & y_1+t/2 \cr -x_1^2-\kappa x_2 &
y_2-x_1 x_2},\eq..
$$
where $\kappa$ is a nonzero constant and $t$ is the deformation parameter
(time). The isotropy group $H_Y$ is
$$
H_Y=\left\{\pmatrix{1&b\cr 0&1}\mid b\in\bfR\right\}.\eq..
$$
Its action on $M$ is generated by
$$
a:=\kappa(y_1-x_2^2)-2 x_1 y_2+x_1^2 x_2.\eq..
$$
This case is exceptional, in that the equations \spcoordeqs do not provide
the coordinate $\ln(w)$ conjugate to $a$, but rather, a constant $w=\kappa$.
Nevertheless, they yield the coordinates $(u,v)$ on the reduced phase space,
$$
u=\kappa^{-1}x_1^2+x_2,\quad v= y_2-x_1 x_2 -\kappa^{-1}x_1^3
\eq..
$$
and we may choose, e.g.
$$
z:={x_1\over\kappa}\eq..
$$
as the remaining canonical coordinate.
The symplectic form is then given by
$$
\ome=du\wedge dv+dz\wedge da\eq..
$$
and the reduced form is
$$
\ome_{\rm red}=du\wedge dv.\eq..
$$
The Hamiltonian $\CalH_I$ on the unreduced manifold is given in terms of
$(x_i,y_i)$ by
$$
\CalH_I(x_1,y_1,x_2,y_2,t)={1\over4}\res_{\la=0}\tr(\la^{-1}\CalN^2(\la)))
={1\over2}\left[(y_2-x_1 x_2)^2-(x_1^2+\kappa x_2)
(y_1+t/2)\right].\eq HI..
$$
The differential operator $\CalD_t$ in \Dopsb defining the isomonodromic
deformation of $\CalD_\la$ corresponding to $\CalH_I$ is obtained by
choosing $\sig=0$ in \Dopsb, giving
$$
\CalD_t={\partial\over\partial t}-\la\,\pmatrix{0&{1\over2}\cr0&0}
-{1\over2}\pmatrix{x_1&x_2\cr\kappa&-x_1}.\eq..
$$
On
the Marsden-Weinstein reduced space corresponding to the invariant level set
$a=a_0$ the reduced Hamiltonian $\tilde\CalH_I$ is given in terms of $(u,v)$ by
$$
\tilde\CalH_I(u,v,t)={v^2\over2}-{\kappa u^3\over 2}-{\kappa ut\over
4}- {a_0 u\over 2}.\eq..
$$
Hamilton's equations for $\tilde\CalH_I$ imply
$$
\ddot{u}={3\kappa\over2}u^2+{\kappa\over4}t +{a_0\over2}.\eq..
$$
This gives the standard form of $P_I$ (cf. \cite{I}) for
$\kappa=4$ and $a_0=0$.

\Untertitel{Painlev\'e II}Let $M=\bfR^4$ again with standard symplectic form
\ddimsym.
The moment map $J_{II}$ into $\gR^*$ is defined by
$$
J_{II}(x_1,x_2,y_1,y_2):=\CalN(\la)=\la^2
\pmatrix{{\kappa\over2}&0\cr0&-{\kappa\over2}} +\la\pmatrix{0 & -\kappa y_1
\cr x_2 & 0} +\pmatrix{x_2 y_1+{t\over2} & -\kappa y_2 \cr x_1 & -x_2
y_1-{t\over2}},\eq..
$$
where $\kappa$ is a nonzero constant and $t$ is the deformation parameter. The
isotropy group $H_Y$ is
$$
H_Y=\left\{\pmatrix{ b&0\cr0&b^{-1}}\mid b\in\bfR^*\right\}.\eq diagiso..
$$
Its action on $M$ is generated by
$$
a:=x_1 y_1+x_2 y_2.\eq..
$$
The spectral Darboux coordinates on the unreduced phase space are given by
$$
u=-x_1/x_2,\ v=-x_2 y_1,\ w=-x_2,\ {\rm and\ }
a=x_1 y_1+x_2 y_2.\eq..
$$
In terms of these, the symplectic form is
$$
\ome=du\wedge dv+d\ln(w)\wedge a.\eq specsymf..
$$
The Hamiltonian $\CalH_{II}$ on $M$ is given by
$$
\CalH_{II}(x_1,y_1,x_2,y_2,t)=
{1\over2\kappa}\res_{\la=0}\tr(\la^{-1}\CalN^2(\la))=
{1\over\kappa}((x_1 y_2)^2+t\,x_2 y_1+t^2/4-\kappa\,x_1 y_2).
\eq HII..
$$
The differential operator $\CalD_t$ in \Dopsb defining the isomonodromic
deformation of $\CalD_\la$ corresponding to $\CalH_{II}$ is
again obtained by setting $\sig=0$ in \Dopsb, giving
$$
D_t={\partial\over\partial t}-\la\,\pmatrix{{1\over2}&0\cr0&-{1\over2}}
-\pmatrix{0&-y_1\cr\kappa^{-1}x_2&0}.\eq..
$$
On the level set
$a=a_0$, the reduced Hamiltonian $\tilde\CalH_{II}$ is expressed
in terms of $(u,v)$ by
$$
\tilde\CalH_{II}(u,v)={1\over\kappa}(v^2-vt+t^2/4-\kappa u(uv-a_0)),\eq..
$$
and the reduced symplectic form $\ome_{\rm red}$ is again
$$
\ome_{\rm red}=du\wedge dv.\eq..
$$
Hamilton's equations for $\tilde\CalH_{II}$ are equivalent to
$$
\ddot{u}=2\,u^3+2\,\kappa^{-1}t\,u+\alpha,\eq PII..
$$
where $\alpha=-\kappa^{-1}(2a_0+1)$. Setting $\kappa=2$, \PII gives the
standard form of $P_{II}$ (cf. \cite{I}).

\Untertitel{Painlev\'e III}Again, $M=\bfR^4$ with symplectic form
\ddimsym. The moment map $J_{III}$ is defined by
$$
\eqalign{
J&_{III}(x_1,x_2,y_1,y_2):=\cr
&\CalN(\la)=\pmatrix{\kappa t&0\cr0&-\kappa t}
-{\smallmatrix{x_1 y_1+x_2 y_2+\mu_1&2\,\left(y_1
y_2-{\mu_1\mu_2\over x_2^{2}} +{\mu_2^2 x_1\over x_2^{3}}\right)
\cr -2\,x_1 x_2&-x_1 y_1-x_2 y_2+\mu_1}\over 2\la}
-\kappa t
{\smallmatrix{
y_1 x_2 + \mu_2 & y_1^2-{\mu_2^2\over x_2^{2}} \cr -x_2^2& -y_1 x_2+\mu_2}
\over2\la^2},}
\eq..
$$
where $\mu_1,\mu_2,\kappa$ are constants, $\kappa\neq 0$ and $t$ is the
deformation parameter.
The isotropy group $H_Y$ is again given by \diagiso and its action on $M$ is
generated by
$$
a:={1\over2}(x_1 y_1+x_2 y_2).\eq..
$$
The spectral Darboux coordinates for this case are
$$
u=-{\kappa t\,x_2\over2\,x_1},\
v=\kappa t{y_1 x_2-\mu_2\over 2\,u^2}+{x_1 y_1+x_2 y_2-\mu_1
\over 2\,u},\ w=-x_1 x_2,\ a={1\over2}(x_1 y_1+x_2 y_2),\eq CoordIII..
$$
with $\ome$ again of the form \specsymf.
On the level set $a=a_0$ the symplectic form $\ome$
projects to
$$
\ome_{\rm red}=du\wedge dv.\eq redsymf..
$$
The relevant Hamiltonian $\CalH_{III}$ on $M$ for this case is given by
$$
\eqalign{
\CalH_{III}=&{1\over t}\res_{\la=0}\tr(\la\,\CalN^2(\la))\cr
=&-2\kappa^2 t y_1 x_2+{2\over t}\left(
{(x_1y_1-x_2y_2)^2\over4}+\mu_1 \mu_2{x_1\over x_2}-\mu_2^2{x_1^2\over x_2^{2}}
\right)+{\mu_1^2\over 2t}.}
\eq HIII..
$$
The differential operator $\CalD_t$ defining the isomonodromic  deformation
equation of $\CalD_\la$ is obtained by setting $\sig=-1/2$ in \Dopsb, giving
$$
\CalD_t={\partial\over\partial t}
-\la\pmatrix{\kappa&0\cr0&-\kappa}
+{1\over 2t}\smallmatrix{x_1 y_1+x_2 y_2+\mu_1 & 2\left(y_1
y_2-{\mu_1\mu_2\over x_2^{2}} +{\mu_2^2 x_1\over x_2^{3}}\right)
\cr -2\,x_1 x_2&-x_1 y_1-x_2 y_2+\mu_1}
-{\kappa\over 2\la}\smallmatrix{
y_1 x_2 + \mu_2 & y_1^2-{\mu_2^2\over x_2^{2}} \cr -x_2^2& -y_1 x_2+\mu_2}.
\eq..
$$
The corresponding Hamiltonian $\tilde\CalH_{III}$ on the reduced phase space is
$$
\tilde\CalH_{III}={2\over t}\left(u^2 v^2-2\,\kappa\,t\,v u^2+\mu_1 u v +
\kappa\,\mu_2\, t v  + (2a_0-\mu_1)\,t\,u\right)+{\mu_1^2\over t}-
2\kappa^2 t\mu_2.
\eq..
$$
Taking into account the explicit $t$-dependence of the coordinates $u,\, v$
defined in \CoordIII, Hamilton's equations for $\tilde\CalH_{III}$ are
equivalent
to the equation of the Painlev\'e transcendent $P_{III}$:
$$
\ddot{u}={(\dot{u})^2\over u}-{\dot{u}\over t}+{1\over t}(\alpha
u^2+\beta)+\gamma u^3 +{\delta\over u},\eq..
$$
where the constants $\alpha,\beta,\gamma,\delta$ are given by
$$
\alpha= -16a_0 +8\mu_1-8\kappa\mu_1-8\kappa,\
\beta=-4\kappa\mu_1 \mu_2,\ \gamma= 16\kappa^2,\
\delta= -4\kappa^2\mu_2^2.\eq..
$$

\Untertitel{Painlev\'e IV}Again, $M=\bfR^4$ with symplectic form
\ddimsym. The moment map $J_{IV}$ is
$$
\eqalign{
J_{IV}&(x_1,x_2,y_1,y_2):=\cr
&\CalN(\la)=\la\pmatrix{1&0\cr0&-1}+\pmatrix{0&{2}y_1\cr-x_1&0}
+{1\over 2(\la-t)}\pmatrix{-x_2 y_2 +\mu & -y_2^2+\mu^2 x_2^{-2}\cr x_2^2&x_2
y_2+\mu},} \eq..
$$
where $\mu$ is constant and $t$ is the deformation parameter. The isotropy
group $H_Y$ is again given by \diagiso and its action on $M$ is generated by
$$
a=x_1 y_1+{x_2 y_2\over2}.\eq..
$$
The spectral Darboux coordinates for this case are
$$
u={x_2^2\over 2 x_1},\ v={x_2 y_2+\mu\over 2 u},\ a=x_1 y_1+{x_2 y_2\over2},
\ w=x_1,\eq..
$$
with $\ome$ again of the form \specsymf.
The unreduced Hamiltonian $\CalH_{IV}$ on $M$ leading to the relevant
isomonodromic deformation equation corresponding to $P_{IV}$ is given by
$$
\CalH_{IV}={1\over2}\res_{\la=t}\tr(\CalN(\la)^2)=-t\, x_2 y_2+x_2^2 y_1
+{1\over2}x_1 y_2^2-{\mu^2\over2}{ x_1\over x_2^2}.\eq HIV..
$$
The corresponding isomonodromic deformation operator $\CalD_t$ is obtained by
setting $\sig=-1$ in \Dopsb, giving
$$
\CalD_t={\partial\over\partial t}+
{\pmatrix{-x_2 y_2 +\mu & -y_2^2+\mu^2 x_2^{-2}\cr x_2^2&x_2
y_2+\mu}\over 2(\la-t)}.\eq..
$$
Fixing the level set $a=a_0$ and
quotienting by the $H_Y$-action, the reduced symplectic form is again of the
form \redsymf, while the reduced Hamiltonian  is
$$
\tilde\CalH_{IV}=-t\,(2 u v-\mu)+(2 a_0+\mu) u-2 u^2 v +u v^2 -\mu v.\eq..
$$
Hamilton's equations for $\tilde\CalH_{IV}$ imply the equation of motion
$$
\ddot{u}={(\dot{u})^2\over 2\,u}+6\,u^3+8\,t\,u^2+2\,(t^2-\alpha)\,u+
{\beta\over u},\eq..
$$
where
$$
\alpha=2a_0+1,\quad \beta=-{\mu^2\over2}.\eq..
$$
The standard normalization for $P_{IV}$ (cf. \cite{I}) is obtained by making
the change\nobreak
$$
\util=2\,u.\eq PIVtransf..
$$

\Untertitel{Painlev\'e V}Again $M=\bfR^4$ with symplectic form
\ddimsym. The moment map $J_{V}$ is defined by
$$
\eqalign{
J_{V}&(x_1,x_2,y_1,y_2):=\cr
&\CalN(\la)=\pmatrix{t&0\cr0&-t}+{\pmatrix{-x_1 y_1-\mu_1& -y_1^2+\mu_1^2
x_1^{-2}\cr x_1^2& x_1 y_1 -\mu_1}\over 2\la}+{\pmatrix{-x_2 y_2-\mu_2&
-y_2^2+\mu_2^2 x_2^{-2}\cr x_2^2& x_2 y_2 -\mu_2}\over 2(\la-1)}}\eq..
$$
where $\mu_1,\mu_2$ are constants and $t$ is the deformation
parameter. The isotropy group $H_Y$ is again given by \diagiso and its action
on
$M$ is generated by
$$
a:={1\over2}(x_1 y_1+x_2 y_2).\eq..
$$
The spectral Darboux coordinates for this case are given by
$$
\display{
u={x_1^2\over x_1^2+x_2^2},\ v={1\over2}\left({x_1 y_1+\mu_1\over u}
+{x_2 y_2+\mu_2\over  u-1}\right),\cr
\ a={1\over 2}(x_1 y_1+x_2 y_2),\ w=x_1^2+x_2^2.}\eq..
$$
The unreduced Hamiltonian $\CalH_{V}$ on $M$ leading to the relevant
isomonodromic deformation equation is given by
$$
\eqalign{
\CalH_V=&{1\over2t}\res_{\la=1}\tr(\CalN(\la)^2)
-{1\over16\,t^2}\left\lbrack\res_{\la=1}\tr(\CalN(\la)^2)+
\res_{\la=0}\tr(\CalN(\la)^2)\right\rbrack^2\cr
=&-{1\over4t}(x_1^2+x_2^2)(y_1^2+y_2^2)+
{1\over4t}\left(\mu_1^2{x_2^2\over x_1^2}+\mu_2^2{x_1^2\over x_2^2}\right)
-x_2 y_2.}\eq HV..
$$
The differential operator $\CalD_t$ defining the corresponding isomonodromic
deformation equation is obtained by setting $\sig=0$ in \Dopsb, giving
$$
\CalD_t={\partial\over\partial t}-\pmatrix{\la&0\cr0&-\la}-{1\over2t}
\pmatrix{0&-y_1^2-y_2^2+\mu_1{y_1\over x_1}+\mu_2{y_2\over x_2}\cr
x_1^2+x_2^2&0}.\eq..
$$
Reducing at the level set $a=a_0$, the reduced Hamiltonian $\tilde\CalH_V$
becomes
$$
\tilde\CalH_V={1\over t}(v^2 u^2-v^2 u -2tvu^2+uv(2t-\mu_1-\mu_2)
+ut(\mu_1+\mu_2+2a_0)+\mu_1 v+\mu_1\mu_2).\eq..
$$
The equations of motion generated by $\tilde\CalH_{V}$ are
$$
\eqalign{
{d^2 u\over dt^2}=&\left({1\over 2u}+{1\over 2(u-1)}\right)\left({du\over dt}
\right)^2-
{1\over t}{du\over dt}-\alpha{u\over t^2(u-1)}-\beta{u-1\over t^2 u}\cr
&-\gamma{u(u-1)\over t}-\delta u(u-1)(2u-1),}\eq..
$$
where
$$
\alpha={\mu_2^2\over 2},\quad\beta=-{\mu_1^2\over 2},\quad\gamma=4a_0+2,\quad
\delta=2. \eq..
$$
The standard form of $P_V$ (cf. \cite{I}) is obtained through the
transformation
$$
\util={u\over u-1}.\eq PVtransf..
$$

\Untertitel{Painlev\'e VI}In this case, we begin with $M=\bfR^6$, with standard
symplectic form
$$
\ome=\sum_{i=1}^3 dx_i\wedge dy_i.\eq..
$$
The moment map $J_{VI}:\bfR^6\ra\gR^*$ is given by
$$
\eqalign{
J_{VI}&(x_1,y_1,x_2,y_2,x_3,y_3):=\cr
&\CalN(\la)={\smallmatrix{-x_1 y_1-\mu_1&-y_1^2+\mu_1^2 x_1^{-2}\cr
x_1^2&x_1 y_1-\mu_1}\over 2\la}+
{\smallmatrix{-x_2 y_2-\mu_2&-y_2^2+\mu_2^2 x_2^{-2}\cr
x_2^2&x_2 y_2-\mu_2}\over 2(\la-1)}+
{\smallmatrix{-x_3 y_3-\mu_3&y_3^2-\mu_3^2 x_3^{-2}\cr
-x_3^2&x_3 y_3-\mu_3}\over 2(\la-t)},}\eq..
$$
where $\mu_1,\mu_2$ and $\mu_3$ are constants and $t$ is the deformation
parameter. In this case $Y=0$ so that the isotropy group $H_Y$ is given by
$$
H_Y=Sl(2).\eq..
$$
The action of $H_Y$ on $M$ is generated by the moment map
$$
J_{Sl(2)}:=\pmatrix{a&b\cr c&-a}\in\grs\grl(2),\eq..
$$
where
$$
\eqalignno{
a&={1\over 2}\sum_{i=1}^3 x_i y_i\eqn PVIDarba.a.\cr
b&=-{1\over 2}(y_1^2+y_2^2-y_3^2)+{\mu_1^2\over 2x_1^2}+{\mu_2^2\over 2x_2^2}
-{\mu_3^2\over2x_3^2}\eqn.b.\cr
c&=-{1\over 2}(x_1^2+x_2^2-x_3^2).\eqn.c.}
$$
The spectral invariants are all invariant under the action of $H_Y$, so
their Hamiltonian flows leave invariant the level set
$$
b=c=0,\eq semireduced..
$$
which we choose as the ``unreduced'' phase space $\Mtil$ for $P_{VI}$. On
$\Mtil$ we may again define spectral Darboux coordinates by
$$
\display{
u={tx_1^2\over w},\quad w=(1+t)x_1^2+tx_2^2-x_3^2\cr
v={1\over2}\left({x_1y_1-\mu_1\over u}+{x_2y_2-\mu_2\over u-1}+
{x_3y_3-\mu_3\over u-t}\right),}\eq CoordVI..
$$
with $a$ defined by \PVIDarba. The symplectic form $\tilde\ome$
on $\Mtil$ is expressed in terms of these coordinates by
$$
\tilde\ome:=\ome\vert_\Mtil=d\ln(w)\wedge da+du\wedge dv.\eq..
$$
The unreduced Hamiltonian $\CalH_{VI}$ on $M$ is given by
$$
\eqalign{
\CalH_{VI}=&{1\over 2}\res_{\la=t}\tr(\CalN(\la)^2)\cr
=&{1\over4t}\left[(x_1y_3+x_3y_1)^2-\mu_1^2{x_3^2\over
x_1^2}-\mu_3^2{x_2^2\over x_3^2}+2\mu_2\mu_3\right]\cr
&+{1\over4(t-1)}\left[(x_2y_3+x_3y_2)^2-\mu_2^2{x_3^2\over x_2^2}
-\mu_3^2{x_2^2\over x_3^2}+2\mu_2\mu_3\right].}\eq HVI..
$$
The differential operator $\CalD_t$ defining the isomonodromic deformation
equation corresponding to $\CalH_{VI}$ is obtained by setting $\sig=-1$ in
\Dopsb, giving
$$
\CalD_t={\partial\over\partial t}+{\pmatrix{-x_3 y_3-\mu_3& y_3^2-\mu_3^2
x_3^{-2}\cr -x_3^2&x_3 y_3-\mu_3}\over2(\la-t)}.\eq..
$$
Restricting to the invariant symplectic manifold $\Mtil$, and reducing at
the level set $a=a_0$ by the 1-parameter group generated on $\Mtil$ by $a$,
the reduced symplectic form is again
$$
\ome_{\rm red}=du\wedge dv\eq..
$$
and the reduced Hamiltonian is
$$
\eqalign{\tilde\CalH_{VI}=&
{1\over t(t-1)}[u(u-1)(u-t)v^2+v(\mu_1(u-1)(u-t)+\mu_2 u(u-t) +\mu_3 u(u-1))\cr
&+\mu_1\mu_2(u-t)+\mu_2\mu_3 u+\mu_1\mu_3(u-1)+(t-u)K],}\eq..
$$
where
$$
K=a_0^2+{1\over4}\left(\sum_{i=1}^3\mu_i\right)^2-{1\over2}\sum_{i=1}^3\mu_i^2.
\eq..
$$
As in the case of $P_{III}$, the explicit $t$-dependence of the coordinate
functions $u, \, v$ defined in \CoordVI must be taken into account when
computing Hamilton's equations. The $t$ derivatives implied by this explicit
dependence are given by
$$
\eqalign{
u_t &= {u(u-1)\over t(t-1)} \cr
v_t &= {-v(2u-1) + a_0 -{1\over 2}(\mu_1+\mu_2+\mu_3) \over t(t-1)}}\eq..
$$

Inserting this into Hamilton's equations for $\tilde\CalH_{VI}$ and eliminating
$v$ gives $P_{VI}$ as the equation of motion:
 $$
\eqalign{
{d^2 u\over dt^2}=&{1\over2}\left({1\over u}+{1\over u-1}+{1\over u-t}\right)
\left(du\over dt\right)^2-\left({1\over t}+{1\over t-1}+{1\over u-t}\right)
{du\over dt}\cr
&+{u(u-1)(u-t)\over t^2(t-1)^2}\left(\alpha +\beta{t\over u^2}+
\gamma {t-1\over (u-1)^2}+\delta{t(t-1)\over(u-t)^2}\right),}\eq..
$$
where
$$
\alpha=2a_0^2+2a_0 +{1\over 2},\quad\beta=-{1\over2}\mu_1^2,\quad
\gamma={1\over2}\mu_2^2,\quad \delta=-{1\over2}\mu_3^2+ {1\over 2}.\eq..
$$

\section.Discussion

With the exception of $P_{IV}$, the isomonodromic deformations derived above
are of the same form as those given in \cite{JM}. The new element here is the
fact that these systems are obtained as nonautonomous Hamiltonian systems
through moment map embeddings into loop algebras, with the Hamiltonians chosen
from the ring of spectral invariants. Moreover, the same ``spectral Darboux
coordinates'' which are used to linearize the flows in the autonomous case lead
here, after suitable reductions, to the canonical coordinates $(u,v)$,
determining the Hamiltonian form of the Painlev\'e equations. The resulting
Hamiltonians coincide essentially with those in \cite{Ok} (after the
transformations \PIVtransf and \PVtransf).

 Prior to reduction, all the Hamiltonians \HI, \HII, \HIII, \HIV, \HV and \HVI
have the general form
$$
\CalH={1\over2}\sum_{i,j=1}^m g^{ij}(y_i+A_i(x))(y_j+A_j(x))+V(x),\eq..
$$
where $m=2$ for $P_I-P_V$ and $m=3$ for $P_{VI}$, for suitably defined
symmetric
contravariant tensor, covector and scalar fields, $g^{ij}$, $A_i$, $v$,
respectively. If the tensors $g^{ij}$ were all non-singular, we could interpret
these as simple magneto-mechanical systems with metric $g_{ij}$ given by the
covariant  tensor inverse to $g^{ij}$, the $A_i$'s interpreted as vector
potentials and the  $V(x)$ as scalar potentials. This is indeed the case for
$P_V$ and $P_{VI}$, although the further symplectic constraints \semireduced
must
be added in the case $P_{VI}$. For $P_V$, Eq.~\HV gives the following metric
tensor, vector potential and scalar fields:
$$
\display{
ds^2=\sum_{i,j=1}^2g_{ij} dx_i dx_j=-{2\,t\over x_1^2+x_2^2}(dx_1^2+dx_2^2)\cr
A=\sum_{i=1}^2A_i dx_i={2\,t\,x_2\over x_1^2+x_2^2}dx_2\cr
V(x)={1\over4\,t}\left(\mu_1^2{x_2^2\over x_1^2}+\mu_2^2{x_1^2\over
x_2^2}\right) +{t\,x_2^2\over x_1^2+x_2^2}+C.}\eq..
$$
For $P_{I}-P_{IV}$ the tensors $g^{ij}$ are singular.

The moment map embeddings characterizing the orbits and choice of
Hamiltonians for the cases $P_V$ and $P_{VI}$ are well understood within the
broader framework \cite{AHH3, HW} of moment map embeddings in rational
$\Ad^*$~orbits of loop algebras. For the other cases, the moment maps and
spectral invariant Hamiltonians have here been constructed case by case,
chosen to suit the known singularity structure of the corresponding
isomonodromic deformation equations. A general formulation, based on moment
maps to loop algebras, allowing higher order poles at $\la=\infty$ as well as
at finite points and a characterization of those spectral invariant
Hamiltonians which generate isomonodromic deformations in the general case has
yet
to be developed.

\references
AHH1&Adams, M.R., Harnad, J. and Hurtubise, J.,
Isospectral Hamiltonian Flows in
Finite and Infinite Dimensions II. Integration of Flows. {\sl Commun. Math.
Phys.}  {\bf 134} (1990), 555--585.

AHH2&Adams, M.R., Harnad, J. and Hurtubise, J.,
Dual Moment Maps into Loop Algebras. {\sl Lett. Math. Phys.} {\bf 20} (1990),
299--308.

AHH3&Adams, M.R., Harnad, J. and Hurtubise, J.,
Darboux Coordinates and Liouville-Arnold Integration in Loop Algebras,
{\sl Commun. Math. Phys.} (1993, in press);
Coadjoint Orbits, Spectral Curves and Darboux Coordinates,
in: ``The Geometry of Hamiltonian Systems'', ed. T. Ratiu,
Publ. MSRI Springer-Verlag, New York (1991);
Liouville Generating Function for Isospectral Hamiltonian
Flow in Loop Algebras, in:  ``Integrable and
Superintergrable Systems'', ed. B. Kuperschmidt, World Scientific,
Singapore (1990).

AHP&Adams, M.R., Harnad, J. and Previato, E.,
Isospectral Hamiltonian Flows in Finite and Infinite Dimensions I.
Generalized Moser Systems and Moment Maps into Loop Algebras.
{\sl Commun. Math. Phys.} {\bf 117} (1988), 451--500.

FN&Flaschka, H. and Newell, A.C., Monodromy- and spectrum preserving
deformations I.{\sl Commun. Math. Phys.} {\bf 76} (1980), 67--116.

H&Harnad, J., Dual Isomonodromic Deformations and Moment Maps into Loop
Algebras. Preprint CRM-{\bf 1844} (1993).

HTW&Harnad, J., Tracy, C.A., Widom, H., Hamiltonian Structure of Equations
Appearing in Random Matrices. Preprint CRM-{\bf 1846} (1993).

HW&Harnad, J., Wisse, A.,  Moment Maps to Loop Algebras, Classical $R$--Matrix
and Integrable Systems, in: ``Quantum Groups, Integrable Models and Statistical
Systems''  (Proceedings of the 1992 NSERC-CAP Summer Institute
in Theoretical Physics, Kingston, Canada, July 1992), ed. J.~Letourneux and
 L.~Vinet, World Scientific, Singapore (1993, in press).

I&Ince, E.L., ``Ordinary Differential Equations'', Dover, New York, 1956.

JM&Jimbo, M., Miwa, T., Monodromy Preserving Deformation of Linear Ordinary
Differential Equations with Rational Coefficients. {\sl Physica 2D} (1981),
407--448.

Ok&Okamoto, K., On the $\tau$-function of the Painlev\'e equations, {\sl
Physica
D} {\bf 2} (1981) 525--535;
The Painlev\'e equations and the Dynkin Diagrams,
in: ``Painlev\'e Transcendents. Their Asymptotics and Physical Applications'',
ed. P.~Winternitz and D.~Levi, Plenum Press, N.Y., {\sl NATO ASI Series B}
Vol. {\bf 278} (1992) 299-313.

S&Semenov-Tian-Shansky, M.A., What is a classical $R$-matrix,
{\sl Funct. Anal. Appl.} {\bf 17} (1983) 259--272.

\endreferences

\end